\documentclass[preprint,review,12pt]{elsarticle}

\usepackage{amssymb}

\journal{Solid State Communications}

\begin{document}

\begin{frontmatter}



\title{Negative-$\mu$ regime in the ac magnetic response of superconductor nanoshells}


\author[a,b]{V. N. Gladilin}
\author[a]{J. Tempere}
\author[a]{J. T. Devreese}
\author[b]{V. V. Moshchalkov}
\address[a]{TQC -- Theory of Quantum and Complex Systems, Universiteit
Antwerpen, Universiteitsplein 1, B-2610 Antwerpen, Belgium}
\address[b]{INPAC -- Institute for Nanoscale Physics and Chemistry,
Katholieke Universiteit Leuven, Celestijnenlaan 200D, B--3001
Leuven, Belgium}



\begin{abstract}
The time-dependent Ginzburg-Landau formalism is applied to analyze
the vortex states and vortex dynamics in superconducting spherical
nanoshells, subjected to mutually perpendicular strong dc and weak
ac magnetic fields. We demonstrate that nonuniformity of the shell
thickness can dramatically affect the ac magnetic response of a 3D
array of superconducting nanoshells. Remarkably, this response is
strongly influenced not only by the relevant geometric and material
parameters and the ac-field frequency but also by the magnitude of
the applied dc field: by changing this field the real part of the
effective ac magnetic permeability can be tuned from positive values
significantly larger than one down to negative values.
\end{abstract}

\begin{keyword}

A. Superconductors; A. Nanoshells; D. Vortices; D. Magnetic
permeability

\end{keyword}

\end{frontmatter}



\section{Introduction}

Over the last decade, advances in nanotechnology have inspired
experimental and theoretical studies of vortex behaviour in
superconductors with curvilinear surfaces: hollow cylinders and
curved
stripes~\cite{liu01,wang05,lu10,chen10,Sabatino2011,Fomin2012},
ferromagnetic–superconducting core–shell structures\cite{Mueller11},
superconducting spheres~\cite{Romaguera07,Shevtsova08,Xu08,Xu09} and
spherical nanoshells~\cite{du04a,du04b,Gladilin08,Tempere2009},
including those with a magnetic dipole
inside\cite{Doria07,Cabral10}. Nanoshells are hybrid nanostructures,
which typically consist of a dielectric core (usually a silicon
oxide nanograin), coated with a thin layer of
metal~\cite{averitt97,oldenburg98}. When the metal in its bulk form
is a superconductor, the nanoshell below the critical temperature
will also exhibit superconductivity in the thin shell around the
insulating core. It has been shown that those nanoshells allow the
coexistence of a Meissner state and a vortex state in equilibrium on
one and the same superconducting layer~\cite{Tempere2009}. In
superconducting spherical nanoshells the surface curvature and the
applied homogeneous magnetic field lead to a Magnus-Lorentz force,
which pushes the vortices and antivortices\cite{note1} towards the
opposite poles of the shell. The same effect can be expected in the
absence of any external magnetic field for spherical
superconductor-ferromagnet nanoshells with uniformly magnetized
core. This can be considered as an effective pinning of vortices and
antivortices at the poles, which strongly affects both the
equilibrium distributions of vortices and their
dynamics~\cite{du04a,du04b,Gladilin08,Tempere2009}.

In this Communication the vortex dynamics in superconducting
nanoshells is investigated for the case when in addition to a
(strong) dc magnetic field, which provides vortex pinning, also a
weak ac magnetic field is present. In particular, we analyze the ac
magnetic response of nanoshells with nonuniform thickness of the
superconducting shell. We show that the effective ac magnetic
permeability of nanoshells, arranged in a 3D array, is very
sensitive to the applied dc magnetic field. For one and the same
array of nanoshells,  the real part of the ac magnetic permeability
at a given frequency can be tuned from relatively large
(``superparamagnetic'') positive values at relatively high dc fields
to negative values at lower dc fields.

\section{Model}

\begin{figure}
\centering
\includegraphics*[width=7.5 cm]{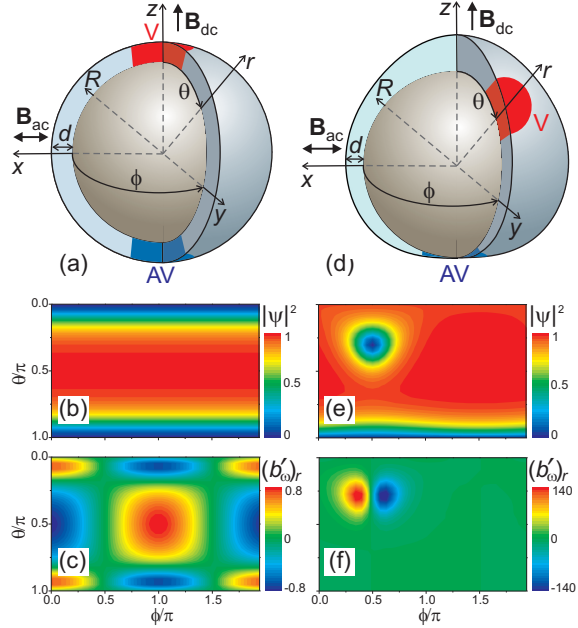}
\caption{(Color online)(a) Sketch of a nanoshell with uniform
thickness $d$. The red and blue shadow schematically show a vortex
and an antivortex, respectively. (b) Distribution on the order
parameter in a nanoshell with $R=2$, $d=0.35$, $\kappa=0.35$ for
$L_x=L_y=L_z=5.2$, $B_{\rm dc}=0.6$ and $B_{\rm ac}=0$. (c)
Distribution of the radial component of the normalized magnetic
field ${\bf b}^\prime_\omega$ in a nanoshell with $R=2$, $d=0.35$,
$\kappa=0.35$ for $L_x=L_y=L_z=5.2$, $B_{\rm dc}=0.6$,
$B_{\omega}=0.003$, $\omega=0.003$. Panels (d) to (f): same as in
panels (a) to (c), respectively, but for the case of the nonuniform
shell thickness described by Eq.~(\ref{thick}) with $d_0=0.35$,
$p=3$ and $q=5$. \label{Fig1}}
\end{figure}

A sketch of a spherical nanoshell with thickness $d$ and radius $R$
is shown in Fig.~\ref{Fig1}(a). A vortex-antivortex pair is induced
in the nanoshell by a homogeneous dc magnetic field ${\bf B}_{\rm
dc}$ parallel to the $z$-axis. An additional oscillating magnetic
field ${\bf B}_{\rm ac}={\bf e}_x B_\omega \cos{(\omega t)}$ with
frequency $\omega$ and amplitude $B_\omega\ll B_{\rm dc}$  is
applied along the $x$-axis. In our calculations we use spherical
coordinates $r$, $\theta$, $\phi$, where $\theta=0$ corresponds to
the positive direction of the $z$-axis in the Cartesian co-ordinate
frame. We consider a spatially periodic 3D rectangular lattice of
identical nanoshells, which are coupled to each other only
electromagnetically. The unit cell of the lattice contains one
nanoshell in the unit-cell center and has sizes $L_x,\ L_y,\ L_z>
2R$, so that the edges of the unit cell do not intersect with the
nanoshell. The amplitude of the ac field is kept relatively small
($B_\omega<10^{-2}$) so that the response of nanoshells to this
field is linear.

To describe the vortex dynamics in nanoshells, we exploit the
time-dependent Ginzburg-Landau (TDGL) approach. Like
in~\cite{Silhanek11}, the relevant quantities are made dimensionless
by expressing lengths in units of $\sqrt{2}\xi$, time in units of
$\pi\hbar/[4k_B(T_c-T)] \approx 11.6\tau_{GL}$, magnetic field in
units of $\Phi_0/(4\pi \xi^2)=H_{c2}/2$, current density in units of
$\Phi_0/(2\sqrt{2}\pi \mu_0 \lambda^2 \xi)$, and scalar potential in
units of $2k_B(T_c-T)/(\pi e)$. Here, $\Phi_0=\pi\hbar/e$ is the
magnetic flux quantum, $\mu_0$ is the vacuum permeability, $\lambda$
is the penetration depth, $\xi$ is the coherence length, $\tau_{GL}$
is the Ginzburg-Landau time, and $H_{c2}$ is the second critical
field.

We consider the case where the thickness of a superconducting shell
is significantly smaller than the coherence length $\xi$, so that
variations of the order-parameter magnitude across the shell as well
as currents in the radial direction are negligible. In general, the
thickness $d$ is assumed to be nonuniform. In this case, using the
aforedescribed units the TDGL equation for the order parameter
$\psi$~\cite{Chapman96}, adapted to the case of spherical geometry,
takes the form
\begin{eqnarray}
\left(\frac{\partial }{\partial t }+i\varphi\right)\psi =2\psi
(1-|\psi |^{2})\nonumber
\\
 +\frac{1}{d}\left(\frac{1}{R\sin{\theta}}  \frac{\partial
}{\partial \phi}-iA_\phi \right)d \left(\frac{1}{R\sin{\theta}}
\frac{\partial }{\partial \phi}-iA_\phi \right)\psi \nonumber
\\
 +\frac{1}{d}\left(\frac{1}{R}  \frac{\partial }{\partial \theta}
+\frac{\cot{\theta}}{R}-iA_\theta \right)d \left(\frac{1}{R}
\frac{\partial }{\partial \theta}-iA_\theta \right)\psi ,
\label{GL1dimless}
\end{eqnarray}
where $\varphi$ is the scalar potential and ${\bf A}$ is the vector
potential.

The scalar-potential distribution is determined from the condition
\begin{eqnarray}
\nabla \cdot {\bf j}=0,  \label{div}
\end{eqnarray}
which reflects the continuity of currents in a superconducting
shell. The total current density ${\bf j}$ in the shell is given by
the sum of the normal and superconducting components:
\begin{eqnarray}
{\bf j}={\bf j}_n+{\bf j}_s,  \label{j}
\end{eqnarray}
\begin{eqnarray}
{\bf j}_n=-\frac{\sigma}{2}\left(\nabla\varphi +\frac{\partial
\bf{A}}{\partial t }\right),  \label{jn}
\end{eqnarray}
\begin{eqnarray}
{\bf j}_s={\rm Im}\left(\psi^*\nabla\psi \right)- {\bf A}|\psi|^2,
\label{js}
\end{eqnarray}
where $\sigma$ is the normal-state conductivity, which is taken as
$\sigma=1/12$ in our units (see Ref.~\cite{kato91}). The vector
potential ${\bf A}$, for which we choose the gauge $\nabla \cdot
{\bf A}=0$, can be represented as
\begin{eqnarray}
{\bf A }={\bf A }_{\rm dc} +{\bf A }_{\rm ac} +{\bf A }_{\rm s}.
\label{atot}
\end{eqnarray}
Here ${\bf A }_{\rm dc}$ denotes the vector potential corresponding
to the dc magnetic field ${\bf B}_{\rm dc}$, while ${\bf A }_{\rm
ac}$ is related to the ac magnetic field. The vector potential ${\bf
A }_{\rm s}$ describes the magnetic fields, induced by the currents
${\bf j}$, which flow in the superconducting shells:
\begin{eqnarray}
{\bf A }_{\rm s} ({\bf r})=\frac{1}{2\pi\kappa^2}\int  d^3 r^\prime
\frac{{\bf j}({\bf r}^\prime)}{|{\bf r}-{\bf r}^\prime |},
\label{a1}
\end{eqnarray}
where $\kappa=\lambda/\xi$ is the Ginzburg-Landau parameter and
integration is performed over the volume of the superconducting
shells. The results, described below, are obtained for a fixed value
of the Ginzburg-Landau parameter ($\kappa = 0.35$).

When solving numerically the TDGL equation~(\ref{GL1dimless}), the
gauge invariance of the discretized expressions is preserved by
introducing link variables following the method of
Refs.~\cite{kato93,gropp96}. Two-dimensional grids, used in our
calculations, typically have $\gtrsim 30R$ equally spaced nodes in
the $\theta $-interval from 0 to $\pi $ and $\gtrsim 45R$ equally
spaced nodes in the $\phi$-interval from 0 to $2\pi $, so that the
distance between the neighboring nodes does not exceed $0.2\xi$
(recall that the dimensionless parameter $R$ is given by the ratio
of the nanoshell radius to $\sqrt{2}\xi$). The step $h_{t}$ of the
time variable $t$ is automatically adapted in the course of
calculation. This adaptation is aimed to minimize the number of
steps in $t$ and -- at the same time -- to keep the solving
procedure accurate: the value of $h_{t}$ is chosen as large as
possible provided that for $t_{\rm new}=t_{\rm old}+h_t$ the
results, calculated with the steps $h_{t}$ and $h_{t}/2$, are
practically the same. Typically, the step $h_{t}$ is $\sim 10^{-5}$
to $\sim 10^{-3}$ depending on a specific distribution of the order
parameter. For momentary distributions of the order parameter and
the vector potential ${\bf A}$, an iteration procedure is used to
determine from Eq.~(\ref{div}) the corresponding distribution of the
scalar potential $\varphi$ with a relative accuracy not worse than
$10^{-4}$. The (time-dependent) vector potential ${\bf A}$ and the
corresponding link variables are calculated using Eq.~(\ref{a1}),
where the contributions of $\sim 10^4$ neighboring nanoshells are
taken into account.

According to Ref.~\cite{pendry99}, the effective relative magnetic
permeability, which corresponds to the response of a nanoshell
lattice under consideration to the ac magnetic field ${\bf B}_{\rm
ac}={\bf e}_x B_\omega \cos{\omega t}$, can be expressed as
\begin{eqnarray}
\mu_\omega=\frac{B_\omega}{B_\omega+\langle B^{(c)}_{{\rm
s}\omega}\rangle_x+i\langle B^{(s)}_{{\rm s}\omega}\rangle_x},
\label{mu1}
\end{eqnarray}
where the numerator originates from the ac $B_x$-field, averaged
over a unit-cell face, normal to the direction of the applied ac
field, while the denominator is determined by the total ac
$H_x$-field, averaged over a unit-cell edge parallel to the applied
ac field, times $\mu_0$. Here the quantities
\begin{eqnarray}
{\bf B }^{(c)}_{{\rm s}\omega}
=\frac{\omega}{\pi}\int\limits_t^{t+2\pi/\omega} dt^\prime {\bf B
}_{\rm s}(t^\prime)\cos{(\omega t^\prime)} \label{bs1}
\end{eqnarray}
and
\begin{eqnarray}
{\bf B }^{(s)}_{{\rm s}\omega}
=\frac{\omega}{\pi}\int\limits_t^{t+2\pi/\omega} dt^\prime {\bf B
}_{\rm s}(t^\prime)\sin{(\omega t^\prime)}, \label{bs2}
\end{eqnarray}
with ${\bf B }_{\rm s}=\nabla\times {\bf A }_{\rm s}$, describe the
in-phase and out-of-phase magnetic field, respectively, induced by
the superconducting nanoshells at frequency $\omega$. As seen from
Eq.~(\ref{a1}), these fields are inversely proportional to
$\kappa^2$. By introducing the normalized complex field ${\bf b
}_{\omega}$ with the real part ${\bf b }^\prime_{\omega}={\bf B
}^{(c)}_{{\rm s}\omega}/B_\omega$ and the imaginary part ${\bf b
}^{\prime\prime}_{\omega}={\bf B }^{(s)}_{{\rm s}\omega}/B_\omega$,
equation~(\ref{mu1}) simplifies into
\begin{eqnarray}
\mu_\omega=\left(1+\langle b_{\omega}\rangle_x\right)^{-1}.
\label{mu2}
\end{eqnarray}

\section{Results and discussion}

\begin{figure}
\centering
\includegraphics*[width=7. cm]{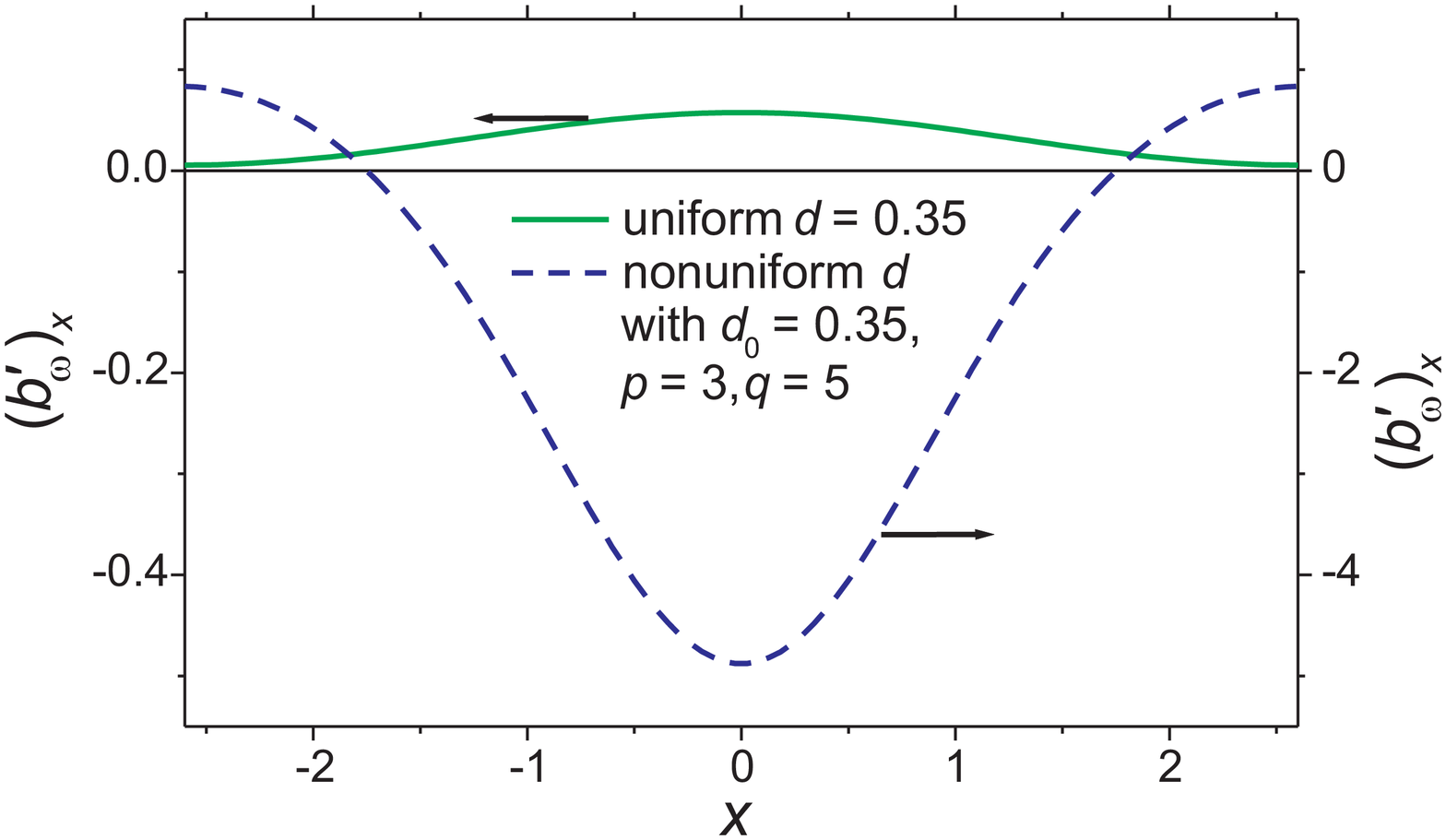}
\caption{(Color online) Distribution of the $x$-component of the
normalized magnetic field ${\bf b}^\prime_\omega$ along a unit-cell
edge, parallel to the $x$-axis, for nanoshells of uniform thickness
$d=0.35$ (solid line) and nonuniform thickness described by
Eq.~(\ref{thick}) with $d_0=0.35$, $p=3$ and $q=5$ (dashed line).
The calculations correspond to $R=2$, $\kappa=0.35$,
$L_x=L_y=L_z=5.2$, $B_{\rm dc}=0.6$, $B_{\omega}=0.003$, and
$\omega=0.003$. \label{Fig2}}
\end{figure}

In Fig.~\ref{Fig1}(b), we plot the equilibrium distribution of the
order parameter in a nanoshell with radius $R=2$ and uniform
thickness $d=0.35$ at $B_{\rm dc}=0.6$ and $B_{\rm ac}=0$. This
distribution corresponds to a single vortex-antivortex pair with the
vortex (antivortex) pinned by the Magnus-Lorentz force to the
northern (southern) pole of the nanoshell. In response to an ac
magnetic field with $B_{\omega}=0.003$, $\omega=0.003$, applied to a
cubic lattice of those nanoshells with $L_x=L_y=L_z=5.2$, the
nanoshells induce their own magnetic field. The radial component of
the corresponding normalized field ${\bf b}^\prime_\omega$ in a
nanoshell is shown in Fig.~\ref{Fig1}(c). In the equatorial regions
of the nanoshell, the magnitude and the sign of the field
$({b}^\prime_\omega)_r $ are mainly determined by the diamagnetic
Meissner currents, induced in the nanoshell by the applied field
${\bf B}_{\rm ac}$. The distribution of $({b}^\prime_\omega)_r$ near
the poles is governed by the oscillatory motion of the vortex and
antivortex. Indeed, the field ${\bf B}_{\rm ac}$, parallel to the
$x$-axis, tends to shift the (anti)vortex from its equilibrium
position on the $z$-axis towards the $x$-axis. Importantly, each
(anti)vortex carries a relatively large magnetic moment, which is
(anti)parallel to the $r$-axis. The local magnetic fields created by
a vortex (antivortex) in the nanoshell and its vicinity are
comparable in magnitude to the relatively strong dc field $B_{\rm
dc}$. Therefore, oscillations of (anti)vortices can produce an
appreciably high ac field. However, due to strong pinning of
(anti)vortices to the nanoshell poles, the oscillation amplitude of
(anti)vortices in the nanoshell lattice under consideration is
rather small. As a result, the normalized magnetic field
$({b}^\prime_\omega)_x$ on the unit-cell edge parallel to the
$x$-axis (see the solid curve in Fig.~\ref{Fig2}) is actually
dominated by the contribution of the Meissner currents.
Consequently, the  effective magnetic permeability of the nanoshell
lattice $\mu_\omega^\prime \approx 0.915$, determined by
Eq.~(\ref{mu2}), corresponds to a diamagnetic medium.

For nanoshells with a nonuniform thickness, the situation can become
significantly different from that described above. Here we restrict
ourselves to the case when the shell thickness $d$ varies according
to the formula
\begin{eqnarray}
d=d_0\left\{\begin{array}{c}
       \left(1+\cos^p{\theta}\right) \quad {\rm for}\ \theta \leq \pi/2\\
       \left(1-|\cos{\theta}|^q\right)\quad {\rm for}\ \theta>\pi/2
     \end{array}\right.
 \label{thick}
\end{eqnarray}
with $p,q>2$, as it is schematically shown in Fig.~\ref{Fig1}(d).
Those variations of $d$ enhance the anivortex pinning at the
southern pole of a nanoshell and, at the same time, tend to expel
the vortex from the northern pole. If these variations are
sufficiently large, the equilibrium position of the vortex in an
isolated nanoshell corresponds to a ring-like potential well at
$\theta\neq 0$. Since this potential well has perfect $C_\infty$
symmetry, the vortex can easily move within the well (in the $\phi$
direction) when applying even a very weak additional force, caused
by the field ${\bf B}_{\rm ac}$. In a rectangular nanoshell lattice
the $C_\infty$ symmetry of the potential well for a vortex is
somewhat violated, mainly due to the magnetic field of the Meissner
currents induced by the field ${\bf B}_{\rm dc}$ in the neighboring
nanoshells. In such a lattice, subjected to a weak ac magnetic field
parallel to the $x$-axis, the vortex typically oscillates around
$\phi=\pi/2$, as illustrated by Figs.~\ref{Fig1}(e) and (f), or
around $\phi=3\pi/2$. At relatively low frequencies $\omega$ the
amplitude of the vortex oscillations can be rather large, and the
radial ac magnetic field induced by these oscillations appears by
orders of magnitude stronger than the applied field $B_\omega$  [see
Fig.~\ref{Fig1}(f)]. The ac field, induced by the nanoshells on the
unit-cell edge, parallel to the $x$-axis, is fully dominated by the
contribution of the oscillating vortices. In the central part of the
unit-cell edge the $x$-component of this field is opposite to the
applied ac field and significantly exceeds it
($|(b_\omega^\prime)_x|>1$; see the dashed curve in
Fig.~\ref{Fig2}). so that the total normalized field
$1+(b_\omega^\prime)_x$ becomes negative.

\begin{figure}
\centering
\includegraphics*[width=7. cm]{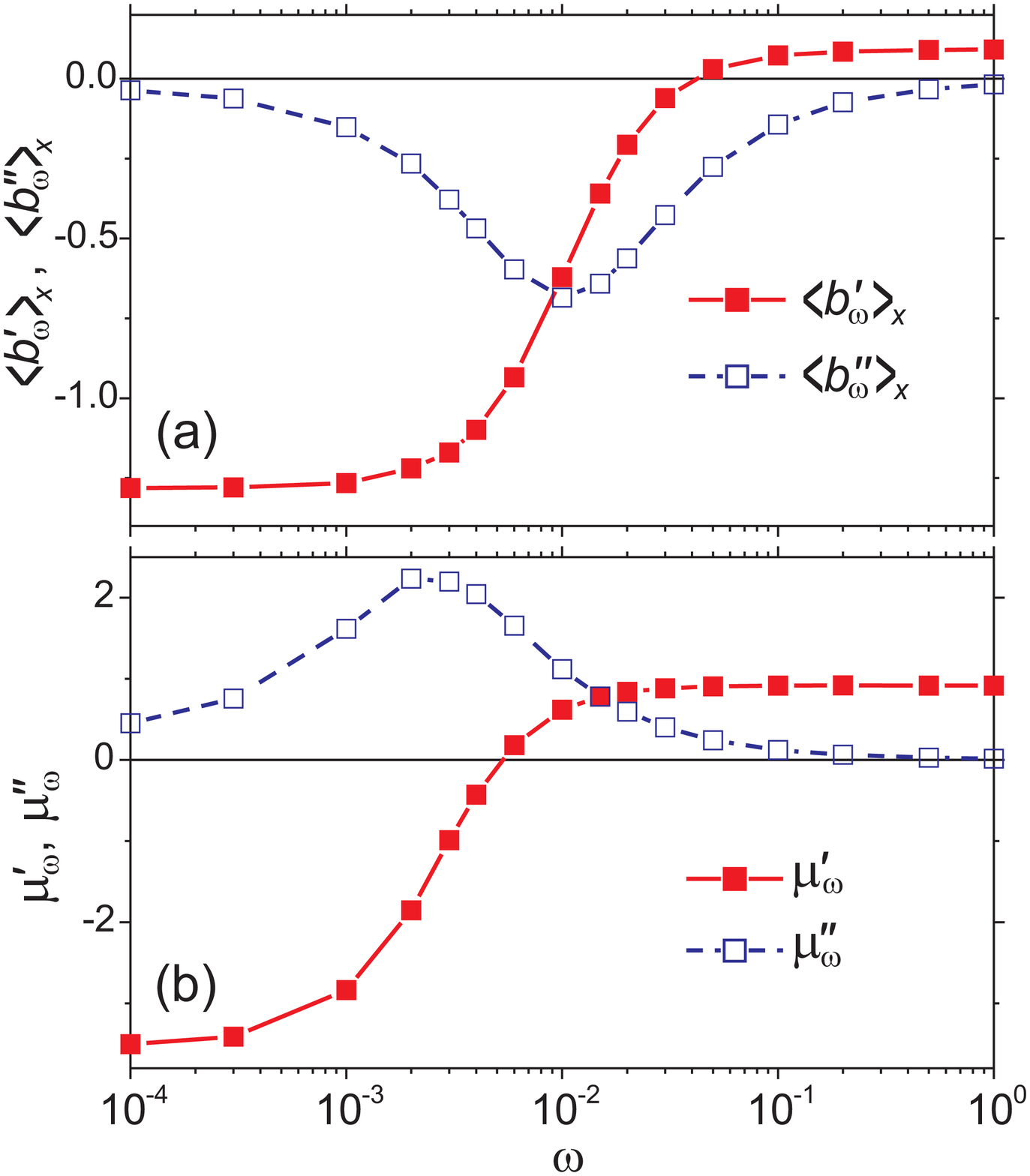}
\caption{(Color online) (a) $x$-components of the normalized
magnetic fields ${\bf b}^\prime_\omega$ and ${\bf
b}^{\prime\prime}_\omega$ averaged over the unit-cell edge, parallel
to the $x$-axis, as a function of the applied ac field frequency.(b)
Real and imaginary parts of the effective magnetic permeability of
the array of nanoshells as a function of the applied ac field
frequency. The calculations correspond to $R=2$, $d_0=0.35$, $p=3$,
$q=5$, $\kappa=0.35$, $L_x=L_y=L_z=5.2$, $B_{\rm dc}=0.8$, and
$B_{\omega}=0.001$. \label{Fig3}}
\end{figure}

In Fig.~\ref{Fig3}(a) the $x$-components of the normalized magnetic
fields ${\bf b}^\prime_\omega$ and ${\bf b}^{\prime\prime}_\omega$
averaged over the unit-cell edge, parallel to the $x$-axis, are
plotted as a function of $\omega$. As seen from Fig.~\ref{Fig3}(a),
at $\omega<0.04$ the averaged field $\langle {
b}^\prime_\omega\rangle_x$ is negative. Moreover, for $\omega<0.004$
its magnitude is larger than 1, so that the real part of the
effective magnetic permeability of the nanoshell lattice
$\mu_\omega^\prime$, defined by Eq.~(\ref{mu2}), appears negative
[see Fig.~\ref{Fig3}(b)]. As further seen from Fig.~\ref{Fig3}, in
the low-frequency range the out-of-phase field component $\langle {
b}^{\prime\prime}_\omega\rangle_x$ and, consequently, the imaginary
part of $\mu_\omega$ rapidly decrease when decreasing $\omega$. This
decrease originates from a reduction of the energy dissipation by
the vortices, which move with smaller average velocities at lower
frequencies. At relatively high frequencies, the amplitude of the
vortex oscillations and their contribution to
$\langle{b}_\omega\rangle_x$ and $\mu_\omega$ diminish with
increasing $\omega$. For $\omega>0.01$ the magnetic response of the
nanoshell lattice is dominated by the contribution of the
dissipationless Meissner currents, induced by the field ${\bf
B}_{\rm ac}$. In this frequency range the effective permeability
$\mu_\omega$ corresponds to a diamagnetic medium, and the imaginary
part of $\mu_\omega$ gradually vanishes with increasing $\omega$.

\begin{figure}
\centering
\includegraphics*[width=7. cm]{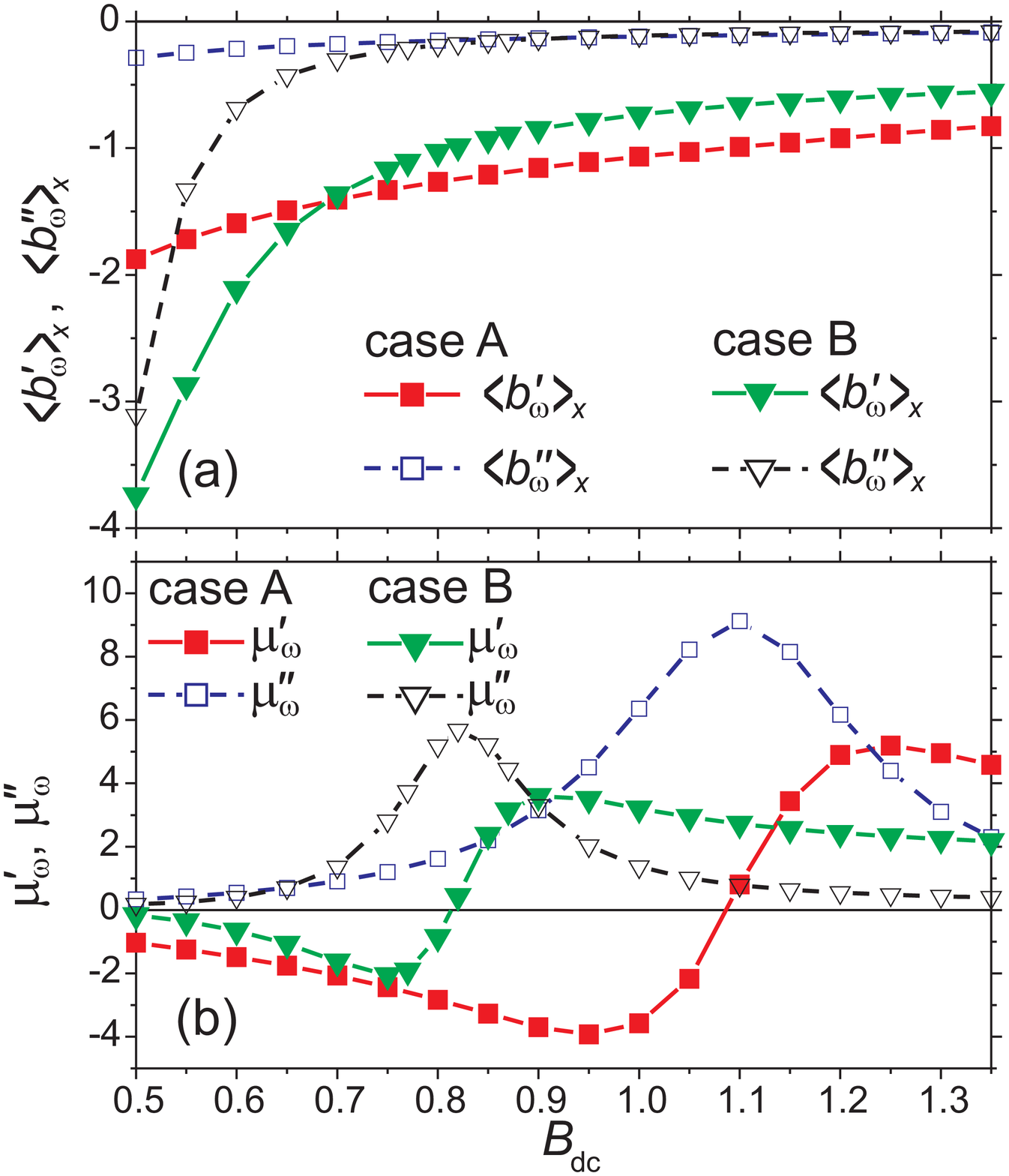}
\caption{(Color online) Averaged $x$-components of the normalized
magnetic fields ${\bf b}^\prime_\omega$ and ${\bf
b}^{\prime\prime}_\omega$ (a) and real and imaginary parts of the
corresponding effective magnetic permeability $\mu_\omega$ (b) as a
function of the applied dc magnetic field. The calculations are
performed for $L_x=L_y=L_z=5.2$, $p=3$, $q=5$, $\omega=0.001$ (case
A) and $L_x=L_y=4.4$, $L_z=6$, $p=q=5$, $\omega=0.003$ (case B).
Other relevant parameters are $R=2$, $d_0=0.35$, $\kappa=0.35$,
$B_{\omega}=0.001$. \label{Fig4}}
\end{figure}

Interestingly, the effective ac magnetic permeability of a nanoshell
array is sensitive not only to the relevant geometric and material
parameters and the ac-field frequency but also to the magnitude of
the applied dc field $B_{\rm dc}$. In this connection, it seems
worth mentioning that in nanoshells with $R\sim 2$ the state with
one vortex-antivortex pair appears (meta)stable in a rather wide
range of $B_{\rm dc}$~\cite{Gladilin08}. As illustrated in
Fig.~\ref{Fig4}, the ac magnetic response of a nanoshell array at a
fixed, relatively low frequency can be efficiently controlled by
simply varying the dc field $B_{\rm dc}$ within this range. Indeed,
the plots in Fig.~\ref{Fig4}(a) show that with increasing $B_{\rm
dc}$ the magnitude of the negative normalized field $\langle {
b}^\prime_\omega\rangle_x$ changes from relatively large values
$|\langle { b}^\prime_\omega\rangle_x|>1$ to values $|\langle {
b}^\prime_\omega\rangle_x|<1$. This behavior can be explained by a
shift of the equilibrium vortex position towards the nanoshell pole
and the corresponding decrease of the angle between the vortex
magnetic moment and the $z$-axis when applying a higher dc field
$B_{\rm dc}$. As seen from Fig.~\ref{Fig4}(a), this leads to an
evolution of the ac magnetic permeability $\mu^\prime_\omega$ from
negative values at lower dc fields $B_{\rm dc}$ to relatively large
positive values, which correspond to (super)paramagnetic ac response
of the nanoshell lattice at higher fields $B_{\rm dc}$.

\section{Conclusions}

Using the time-dependent Ginzburg-Landau approach, we have
investigated vortex states in superconducting spherical nanoshells
with nonuniform thickness, subjected to a homogeneous dc magnetic
field, and analyzed the vortex dynamics in the presence of an
additional weak ac magnetic field, perpendicular to the dc field. It
is shown that by increasing the shell thickness at the pole,
determined by the direction of the dc field, the equilibrium vortex
position can be shifted from this pole to the surrounding ring-like
potential well. In this state, the amplitude of vortex oscillations
in an ac magnetic field appears strongly enhanced, so that the
magnetic response of a nanoshell is fully dominated by the
contribution of the oscillating magnetic moment of the vortex. As
distinct from the case of nanoshells with uniform thickness, which
are typically characterized by a diamagnetic ac response, for a 3D
lattice made of nanoshells with nonuniform thickness the real part
of the effective ac magnetic permeability can take negative values
at sufficiently low frequencies. We have demonstrated that the ac
magnetic response of such a lattice can be efficiently controlled by
the magnitude of the applied dc magnetic field. By varying this dc
field the real part of the effective ac magnetic permeability of the
nanoshell lattice can be tuned in a wide range: from negative values
to relatively large positive (``superparamagnetic'') values.

\section{Acknowledgements}

This work was supported by Methusalem funding by the
Flemish government, the Flemish Science Foundation (FWO-Vl), in
particular FWO projects G.0365.08, G.0370.09N, , G.0119.12N, and
G.0115.12N, the Scientific Research Community project WO.033.09N,
the Belgian Science Policy, and the ESF NES network.








\end{document}